\documentclass{PoS}

\title{Looking at the photoproduction of massive gauge bosons at the LHeC}

\ShortTitle{Looking at gauge bosons photoproduction at the LHeC}

\author{\speaker{Magno MACHADO}\thanks{This work was  partially financed by the Brazilian funding agencies CNPq and FAPERGS and by the French-Brazilian scientific cooperation project CAPES-COFECUB 744/12.}\\
        Instituto de Fisica - Universidade Federal do Rio Grande do Sul\\
	Caixa Postal 15051, CEP 91501-970, Porto Alegre, RS, Brazil \\
        E-mail: \email{magnus@if.ufrgs.br}}

\author{Cristiano BRENNER MARIOTTO\\
        Inatituto de Matem\'atica, Estatistica e Fisica , Universidade Federal do Rio Grande\\
       Avenue It\'alia, km 8, Campus Carreiros. CEP 96203-900, Rio Grande, RS, Brazil\\
        E-mail: \email{cristianomariotto@furg.br}}

\abstract{In this contribution we report on the investigation of the photoproduction of $W^{\pm}$ and $Z^0$ bosons in the planned electron-proton/nucleus collider, the LHeC. The production cross sections and the number of events are provided and theoretical uncertainties are discussed. We also analyze the sensitivity of the LHeC experiment to physics beyond Standard Model by studying the role played by anomalous $WW\gamma$ coupling in the presented process.} 

\FullConference{XXI International Workshop on Deep-Inelastic Scattering and Related Subject -DIS2013,\\
		22-26 April 2013\\
		Marseilles,France}

\begin{document}
\section{Introduction}
Being planned to start in the mid-2020s, the Deep Inelastic Electron-Nucleon Scattering at the LHC (LHeC) machine is a possible extension of the current LHC at CERN, an electron-proton collider \cite{dainton}. It is a convenient way to go beyond the LHC capabilities, exploiting the 7 TeV proton beams which will be produced at the LHC, to drive research on $ep$ and $eA$ physics at some stage during the LHC time. This LHC extension will open a new kinematic window - the $\gamma p$ CM energy can reach up to TeV scale, far beyond the  $\sqrt{s}\gg$  200 GeV at HERA, a very fruitful region for small-$x$ physics and many other physics studies. Despite of great successes of Standard Model (SM) and difficulties to  find new physics such as supersymmetry, which is the most popular scenario, the non abelian self-couplings of $W$, $Z$ and photon remains poorly measured up to now. In this context, the investigation of three gauge boson couplings plays an important role to manifest the non abelian gauge symmetry in standard electroweak theory. Their precision measurement will be the crucial test of the structure of the SM. The inclusive and exclusive production of $W$ and $Z$ at the LHC already provides important tests of SM and beyond. However, the photoproduction channel has the advantage to be much cleaner than the $pp$ collision channels. Here, we investigate the photoproduction of massive gauge bosons at TeV scale and also examine the potential of LHeC collider to probe anomalous $WW\gamma$ coupling. Along these lines, we propose some observables which are sensitive to deviations from SM physics.

In this contribution we summarize the main results obtained in Ref. \cite{R2}. Namely, it is presented the predictions for the photoproduction of massive gauge bosons at future LHeC energies within SM physics. Moreover, we consider physics beyond SM in the case of photoproduction of W bosons analyzing the production rates of $W$ bosons. The sensitivity of LHeC for deviations from SM are investigated and some additional observables are proposed.

\section{Theoretical framework and numerical calculations}

First, we consider the C and P parity conserving effective Lagrangian for two charged W-boson and one photon interaction \cite{hagiwara}. The motivation is to use the $W$ photoproduction cross section as a test of the $WW\gamma$ vertex. In such a case, it is introduced two dimensionless parameters $\kappa$ and $\lambda$ which are related to the magnetic dipole and electric quadrupole moments. Namely, $\mu_W = \frac{e}{2m_W}(1+\kappa+\lambda)$ and $Q_W = -\frac{e}{m_W^2}(\kappa-\lambda)$. In the case of values $\kappa=1$ and $\lambda=0$  Standard Model (SM) is recovered at tree level. We are left with three diagrams for the subprocess $\gamma q_{i}\rightarrow Wq_{j} $ and  only t-channel W exchange graph contributes $WW\gamma$ vertex. The unpolarized differential cross section for the subprocess $\gamma q_{i}\rightarrow Wq_{j} $ can be obtained using helicity amplitudes from summing over the helicities. For the signal we are considering a quark jet and on-shell W with leptonic decay mode $\gamma p\rightarrow W^{\mp}+jet\rightarrow\ell+p_{T}^{miss}+jet $, where $\ell=e,\mu$. In the current mode charged lepton and the quark jet are nicely separated and the signal is prevented from background of SM.

The cross section for the subprocess $ \gamma q_{i} \rightarrow W q_{j}$ is composed by the direct and resolved-photon production, $\hat{\sigma}=\hat{\sigma}_{dir}+\hat{\sigma}_{res}$. The direct-photon contribution is given by \cite{WWg2,WWg3}:
\begin{eqnarray}
\hat{\sigma}_W &=&\sigma_0\{|V_{q_{i}q_{j}}|^{2}\{(|e_{q}|-1)^{2}(1-2\hat{z}+2\hat{z}^{2})
\log({\hat{s}-M_{W}^{2}\over\Lambda^{2}})  -  [(1-2\hat{z}+2\hat{z}^{2}) - 2|e_{q}|(1+\kappa+2\hat{z}^{2})
+{{(1-\kappa)^{2}}\over{4\hat{z}}} \nonumber \\
&- & {{(1+\kappa)^{2}}\over{4}}]
\log{\hat{z}}+  [(2\kappa+{{(1-\kappa)^{2}}\over{16}})
{1\over \hat{z}}
 + ({1\over 2}
 +  {{3(1+|e_{q}|^{2})}\over{2}})\hat{z}
 + (1+\kappa)|e_{q}|-{{(1-\kappa)^{2}}\over{16}} \nonumber \\
& + & {|e_{q}|^{2}\over 2}](1-\hat{z})
-{{\lambda^{2}}\over{4\hat{z}^{2}}}(\hat{z}^{2}-2\hat{z}
\log{\hat{z}}-1)
+ {{\lambda}\over{16\hat{z}}}
(2\kappa+\lambda-2)[(\hat{z}-1)(\hat{z}-9)
+4(\hat{z}+1)\log{\hat{z}}]\}, \nonumber \\
\label{sigmadir}
\end{eqnarray}
where  $\sigma_0 ={{\alpha G_{F}M_{W}^{2}}\over{\sqrt{2}\hat{s}}}$,  $\hat{z}=M_{W}^{2}/\hat{s}$ and $\Lambda^{2}$ is
cut off scale in order to regularize $\hat{u}$-pole of the
collinear singularity for massless quarks. In addition, $\Lambda^2$ is the scale which determines the running of photon structure functions in the resolved part.  The quantity $V_{ij}$ is the Cabibbo-Kobayashi-Maskawa (CKM) matrix and  $e_{q}$ is the quark charge.

The direct part of  cross section then reads as:
\begin{eqnarray}
\sigma_{dir}(\gamma p \rightarrow W^{\pm}X)=\int_{x_p^{m}}^{1}dx_p\sum_{q,\bar{q}}f_{q/p}(x_p,Q^{2})\,\hat{\sigma}_W(\hat{s}),
\end{eqnarray}
where $f_{q/p}$ are the parton distributions functions in the proton, $x_p^{m}=m_W^2/s$ and $\hat{s}=x_ps$

The resolved-photon part of cross section can be calculated
using the usual electroweak formula for $q_{\gamma}q_{p}\to W^{\pm}$ fusion process, $\hat{\sigma}(q_i\bar{q}_j\rightarrow W)=\frac{\sqrt{2}\pi}{3}G_Fm_W^2|V_{ij}|^2\delta (x_ix_js_{\gamma p}-m_W^2)$.
For the photoproduction cross sections one needs parton distribution
functions inside the photon and proton. The photon structure
function $f_{q/\gamma}$ consists of perturbative point-like parts 
and hadron-like parts. Putting all together, the resolved-photon part reads as:
\begin{eqnarray}
\sigma_{res}(\gamma p \rightarrow W^{\pm}X)  =  \frac{\pi\sqrt{2}}{3\,s}G_{F}m_{W}^{2}
|V_{ij}|^{2}\int_{x_{\gamma}^m}^{1}\frac{dx_{\gamma}}{x_{\gamma}}\sum_{q_i,q_j}f_{q_{i}/p}
(\frac{m_{W}^{2}}{xs},Q_{p}) \left[f_{q_{j}/\gamma}(x_{\gamma},
Q_{\gamma}^{2})-\tilde{f}_{q_{j}/\gamma}(x_{\gamma},Q_{\gamma}^{2})\right] \nonumber \\
\end{eqnarray}
where in order to avoid double counting on the leading
logarithmic level, one subtracts the point-like part of photon structure function (photon splitting at large $x$), $\tilde{f}_{q/\gamma}(x,Q_{\gamma}^{2})=\frac{3\alpha e_{q}^{2}}{2\pi}[x^{2}+(1-x)^{2}]\log (Q_{\gamma}^{2}/\Lambda^{2})$. In addition, here $x_{\gamma}^m=m_W^2/s$.

Similar calculation can be done for the $Z$ boson photoproduction. Once again, the cross section in the SM model for the subprocess $ \gamma q \rightarrow Z q$ is composed by the direct and resolved-photon production, $\hat{\sigma}=\hat{\sigma}_{dir}+\hat{\sigma}_{res}$. Explicit expressions for them can be found in Ref. \cite{R2}.

Let us now perform estimates for the LHeC regime using the design with a electron beam having laboratory energy of $E_e=70$ GeV, the center of mass energy will reach $E_{cm}=W_{\gamma p}=1.4$ TeV and a nominal luminosity of order $10^{33}$ cm$^{-2}$s$^{-1}$. One gets $\sigma(\gamma +p\rightarrow W^{\pm}X)\simeq 400$ pb and $\sigma(\gamma +p\rightarrow Z^0X)\simeq 60$ pb considering the SM cross sections.   We have summed the resolved and direct contributions. It is seen that the cross sections are at least one order of magnitude larger than for DESY-HERA machine, $W_{\gamma p }\simeq 300$ GeV. We have checked that our calculations fairly reproduce the DESY-HERA data \cite{R6}, $\sigma_{\mathrm{exp}}(ep\rightarrow eWX)= 1.06\pm 0.17$ pb (combined H1 and ZEUS data). Our result is $\sigma (ep\rightarrow eWX)= 1.3$ pb
 with a typical 15 \% error from the theoretical uncertainties.

In Table \ref{tab1} the photon-proton total cross sections times branching
ratio of $W\rightarrow \mu\nu $ and corresponding number of events
 are shown for SM parameters for W ($\kappa=1$ and $\lambda=0$) and also for $Z^0$ boson with corresponding branching
ratio of $Z^0\rightarrow \mu^+\mu^- $. The number of events has been computed using $N_{ev}=\sigma(e p\rightarrow V+X)BR(V\rightarrow \mu\nu /\mu^+\mu^-){\cal L}$. At this point we consider the acceptance in the leptonic channel as 100\%.   The photoproduction cross section is calculated by convoluting the Weizs\"{a}cker-Williams spectrum with the differential hadronic cross section. Through the calculations proton
structure functions of CTEQ \cite{Pumplin:2002vw} and  photon structure functions
of GRV \cite{grvphoton} have been used
with $Q^{2}=M_{W}^{2}$. The usual electroweak parameters are taken from Ref. \cite{PDG}. We have assumed an integrated
luminosity ${\cal L}$ as 10 fb$^{-1}$ \cite{desreport} in order to compute the number of events, $N_{ev}$. The number of events is large enough to putting forward further analysis as we have units of events per second for $W^{\pm}$.

\begin{table}[t]
\begin{center}
\begin{tabular}{|l|c|c|}
\hline
  $V$  & $\sigma(\gamma p\rightarrow V\,X) \times BR$& $N_{ev}$ \\
\hline
$W^+$ & 24 & $1.2\times 10^{4}$ \\
$W^-$ & 24 & $1.2\times 10^{4}$ \\
$Z^0$ & 2.1 & $1.1\times 10^{3}$ \\
\hline
\end{tabular}
\caption{The photon-proton cross sections times branching ratios
$\sigma(\gamma p\rightarrow W^{\pm}X)\times BR(W^{+}\rightarrow
 \mu\nu)$  and $\sigma(\gamma p\rightarrow Z^0X)\times BR(Z^0\rightarrow
 \mu^+\mu^-)$ in units of pb. The number of events $N_{ev}$ is also presented at integrated luminosity 10 fb$^{-1}$.
\label{tab1}}
\end{center}
\end{table}

Certain properties of the $W$ bosons such as the magnetic dipole and the electric quadrupole moment play a role in the interaction vertex $WW\gamma$, thus processes involving this vertex offer the opportunity to measure such properties. The magnetic dipole moment $\mu_W$
and the electric quadrupole moment $Q_W$ of the $W$ bosons can be written in terms of parameters $\kappa, \,\lambda$, where $\kappa=1$ and $\lambda=0$ are the Standard Model values for those parameters at tree level. In $W$ photoproduction one has a unique scenario to test the anomalous $WW\gamma$ vertex and its $\kappa$ and $\lambda$ parameters. In the photoproduction of $W^{\pm}$ bosons, the direct contribution $\sigma_{dir}$ involves the generalized $WW\gamma$ vertex, and then the expression for $\hat{\sigma}_W(\hat{s}=x_ps)$ in Eq. (\ref{sigmadir}) can be used to investigate deviations from SM physics. An interesting observable is the number of muon plus neutrino events coming from the decay of the $W^+$. This is shown in Table \ref{tab2}, where we assumed  the  luminosity of ${\cal {L}}=10\,$ fb$^{-1}$. The number of $W^+\to\mu\nu$ events is very dependent on the choice of the $\kappa$ and $\lambda$ parameters, and in most scenarios it increases as $\kappa$, $\lambda$ increases/depart from Standard Model.

\begin{table}[t]
\begin{center}
\begin{tabular}{|l|l|c|c|}
\hline
  $\kappa$ & $\lambda$  & $\sigma(\gamma p\rightarrow W^+\,X) \times BR$ [pb] & $N_{ev}$ \\
\hline
 0 & 0 & 16 & $8\times 10^{3}$ \\
 1 & 0 & 24 & $1.2\times 10^{4}$ \\
 2 & 0 & 44 & $2.2\times 10^{4}$ \\
 1 & 1 & 61 & $3.1\times 10^{4}$ \\
 1 & 2 & 172 & $8.5\times 10^{4}$\\
\hline
\end{tabular}
\caption{The number of muon plus neutrino events coming from the $W^+$ decay for distinct choices for the parameters $\kappa$ and $\lambda$ presented at integrated luminosity of 10 fb$^{-1}$.
\label{tab2}}
\end{center}
\end{table}

Finally, in Ref. \cite{R2} we consider different observables which could contribute together to pin down the correct $WW\gamma$ vertex. We introduced the ratio $\sigma (W^{\pm})/\sigma (Z)$ which is less sensitive to the NLO QCD corrections and the $W$-asymmetry observable $A(\kappa, \lambda; \sqrt{s})$ which scans asymmetries in the $W$-photoproduction. The first one has been already proposed a long time in Refs. \cite{WWg2,WWg3}. It was shown that the ratio $\sigma (W^{\pm})/\sigma (Z)$ has high sensitivity to the $\kappa$ and $\lambda$ parameters for LHeC energies \cite{R2}. Therefore, the LHeC collider would be able to pin down the correct values for these parameters and then to determine the magnet dipole and electric quadrupole of the $W$. Moreover, the $W^{+}W^{-}$ asymmetry defined by $A(\kappa, \lambda; \sqrt{s})=\frac{(\sigma_{W^+}-\sigma_{W^-})}{(\sigma_{W^+}+\sigma_{W^-})}$ depends strongly on the $\kappa$ and $\lambda$ parameters, and is therefore a useful observable to help in determine the best scenarios. As a general conclusion about the anomalous coupling, we see that the LHeC has better sensitivity to the parameters $\kappa$ and $\lambda$ compared to DESY-HERA $ep$ collider and it would give complementary information to the LHC collider.

\section{Conclusions}
\label{conc}

As a summary, we have examined the prospects for massive gauge bosons detection at proposed LHeC machine. The photon-proton cross sections have been computed for $W^{\pm}$  and $Z^0$ inclusive production and are of order dozens of picobarns. The number of events is evaluated for the photoproduction cross section assuming an integrated luminosity of 10 fb$^{-1}$ and they are large enough to turn out the measurements feasible. We have investigated also the anomalous $WW\gamma$ coupling using the machine design. We found that kinematic limit to be available at LHeC is somewhat increased relative to the previous DESY-HERA machine. We have tested some sample scenarios beyond SM physics by scanning the values of parameters $\kappa$ and $\lambda$ considering anomalous $WW\gamma$ coupling. In the case of anomalous coupling, the photoproduction process at the LHeC proves to be a powerful tool.

\end{document}